# Implementation of temporal ptychography algorithm, i$^2$PIE, for improved single-beam coherent anti-Stokes Raman Scattering measurements

RUAN VILJOEN[1], PIETER NEETHLING[1,*], DIRK SPANGENBERG[1,2], ALEXANDER HEIDT[2], HANS-MARTIN FREY[2], THOMAS FEURER[2] AND ERICH ROHWER[1]

[1]*Laser Research Institute, Stellenbosch University, Stellenbosch, South Africa.*
[2]*Institute for Applied Physics, University of Bern, Sidlerstrasse 5, 3012 Bern, Switzerland.*
*\*pietern@sun.ac.za*

**Abstract:** We present an improvement on the signal to background of single-beam coherent anti-Stokes Raman scattering (SB-CARS) spectroscopy measurements, for systems employing ultrafast supercontinuum sources based on all-normal dispersion photonic crystal fibers. Improvements to the signal to background arise in the use of a new pulse reconstruction algorithm based on temporal ptychography, i$^2$PIE. A simple SB-CARS strategy is used to measure the spectrum of para-xylene where the supercontinuum pulses used are compressed using multiphoton intrapulse interference phase scan (MIIPS) and for the first time i$^2$PIE using the same single-beam setup. With the i$^2$PIE implementation the signal to background is improved by nearly a factor of 4 in comparison to MIIPS. More notably, the integrated SB-CARS spectral intensity is increased by a factor of 6.5.

## 1. Introduction

Broadband supercontinuum light sources have become valuable tools used for nonlinear microscopy and spectroscopy applications. Properly compressed, low average power supercontinuum laser pulses can efficiently drive nonlinear processes, due to their high peak power, allowing for noninvasive study of biological or chemical systems [1]. One such nonlinear process, Coherent anti-Stokes Raman Scattering (CARS), can probe the vibrational spectrum of molecules. It comes in different guises for a variety of applications from trace chemical detection [2] to imaging brain structures [3], and can be used to expand existing microscopy systems with multimodal functionality [4]. One variation of CARS, which is especially useful in microscopy and spectroscopy, is single-beam CARS (SB-CARS) [5]. SB-CARS employs a simple single beam geometry with a single broadband laser source, capable of simultaneously stimulating and measuring a broad range of molecular vibrations. We have spent the past 3 years developing a SB-CARS apparatus that is the only one of its kind in South Africa and distinctive in the global context.

SB-CARS microscopy requires low average power, high peak intensity, and high repetition rate laser sources. Low average power ensures minimal heating of the sample, high peak intensity results in an increase in generated signal and a high repetition rate decreases scan times. An all normal dispersion photonic crystal fiber (ANDi-PCF) pumped by a stable fs oscillator fulfills all these requirements. The inherent pulse-to-pulse stability of the oscillator is preserved in the generated CARS spectrum, despite the laser pulse undergoing two sequential third order processes, initially spectral broadening in the ANDi-PCF (predominantly self-phase modulation) and then four wave mixing to produce the CARS signal. Furthermore, the phase stability of the ANDi-PCF output allows repeatable and reproducible pulse compression, ensuring temporally optimized pulses [6]. These compressed pulses are considerably lower in energy than those originating from the oscillator, but with a peak power that exceeds the oscillator output. In this way the source fulfills all the ideals of a nonlinear microscopy laser source [7].

Major efforts were made to develop a suitable broadband light source, not only important for SB-CARS but one that can also be employed for other nonlinear imaging modalities. Building on previous work done in our group on supercontinuum generation in fibers [8], an all-normal dispersion photonic crystal fiber (ANDi-PCF), with a femtosecond oscillator as a pump, was integrated into the setup. The dominant driving mechanisms of the supercontinuum generation in ANDi-PCFs are self-phase modulation and optical wave breaking, resulting in a highly coherent and phase stable spectrum [8]. A combination of the generation mechanisms and the dispersive properties of the PCF produce supercontinuum pulses with a complex spectral phase function, leading to temporally broadened pulses. Ideally, a supercontinuum with a flat spectral phase is desired for SB-CARS and nonlinear processes in general, as this ensures the shortest possible pulses.

Unfavorable results from our implementation of an inline characterization and pulse compression scheme, multi photon intrapulse interference phase scan (MIIPS) [9], led to the development of a new ptychography based pulse characterization technique, $i^2$PIE [10], that we show is more suited to our light source. The pulse shaper, a spatial light modulator in 4f-shaper geometry, allows for modulation of the spectral phase, amplitude, and polarization. This enables the implementation of the pulse characterization schemes as well as different SB-CARS strategies in an experimental setup with a single collinear geometry. Interaction between the light source and samples for SB-CARS, MIIPS and $i^2$PIE measurements occur in the focus of the same high numerical aperture microscope objective. In other words, after pulse compression using either characterization technique, compressed pulses are delivered at the sample plane for SB-CARS measurements without changing the experimental setup.

In this article we show, for the first time, the improvement observed in SB-CARS measurements with the implementation of $i^2$PIE in our setup. We compare two Raman spectra obtained from processed SB-CARS measurements, using MIIPS and $i^2$PIE to compress the supercontinuum pulses. We also show our capability of employing another SB-CARS strategy that was previously not possible with low average power due to inadequate pulse characterization and compression using MIIPS.

## 2. Single-beam CARS

In single-beam CARS one spectrally broad pulse simultaneously acts as the pump, Stokes, and probe fields [11]. An increase in the spectral bandwidth increases both the resonant and non-resonant CARS intensities. However, the non-resonant contribution increases much more drastically as a function of bandwidth when compared to the resonant contribution, quickly becoming many orders of magnitude larger than the resonant contribution [12]. Several strategies have been introduced to manage the non-resonant CARS contribution with the aim to extract the resonant signal, which contains the molecular vibrational information. Some of these retrieve the resonant CARS spectrum by using the non-resonant background as a local oscillator to enhance the resonant CARS [5,13,14], while others reduce the non-resonant background [15]. The resonant CARS can also be retrieved by Fourier transforming time resolved SB-CARS measurements [16].

Pulse shaping is ubiquitous among these SB-CARS strategies, as it allows for the control of the amplitude, phase, and polarization of the spectral components of the light source. Each technique tailors the excitation pulses differently to affect the non-resonant and resonant

CARS polarization. In turn, this affects the measured CARS intensity, as it is proportional to the superposition of the non-resonant and resonant polarization. The CARS response due to the tailored pulses is easily predictable as the CARS process is well described [17]. The success of the various strategies relies on the pulses having a flat spectral phase (or as close to flat as possible), as the different spectral components have to be incident on the molecule at the same time (in phase) in order to generate the instantaneous CARS response in a sufficiently broad spectral range. Using a pulse with a complex phase structure will lead to unpredictable results and reduced phase-matching bandwidth, diminishing the ability to extract the resonant CARS spectrum.

We highlight two SB-CARS strategies to (1) illustrate the improvement observed when implementing our new pulse characterization technique, i$^2$PIE, and (2) show that we are able to implement more complex phase shaping strategies to isolate specific resonant Raman excitations of interest.

## 2.1 Narrowband probe with phase shaping

In order to illustrate the improved SB-CARS response we implement a simplified narrowband pulse shaping strategy. We choose a narrow (1.27 nm) spectral slice (indicated by the purple area in figure 1.B) from the high energy side of the supercontinuum to act as the probe, which allows one access to low wavenumbers in the CARS spectrum. The remaining part of the supercontinuum acts as the pump [18].

For this strategy the narrowband probe is modulated with a $\pm \pi / 2$ phase-change, using an SLM in a 4-f shaper geometry, and the resultant CARS spectra are measured for the addition and subtraction of the spectral phase. Spectral features arise due to the interference of the non-resonant and resonant CARS contribution. In the measured CARS spectra this translates to, a dip/peak at the spectral positions which correspond to the Raman line, relative to the spectral position of the probe. With this information one can infer the Raman spectrum by subtracting the two measured spectra from each other and correcting for the spectral amplitude of the driving field:

$$R(\Omega) = \frac{S^-(\Omega) - S^+(\Omega)}{\sqrt{S^-(\Omega) + S^+(\Omega)}} \left[ \int_0^\infty E(\omega') E^*(\omega' - \Omega) d\omega' \right]^{-1}. \qquad (1)$$

Here $E(\omega)$ is the broadband supercontinuum field and the integral term represents the second order polarization driving the molecular vibrations. $S^\pm$ are the measured intensity spectra corresponding to the respective phase changes and have been shifted relative to the probe central frequency ($\Omega = \omega_{anti-Stokes} - \omega_{probe}$). The result, $R(\Omega)$, corresponds to the spontaneous Raman spectrum, and should have the same relative peak heights.

## 2.3 Spectral focusing with binary phase shaping

Beyond simply measuring the Raman spectrum with SB-CARS, other groups have developed extended phase shaping schemes that are capable of isolating chosen resonant excitations and focusing them by applying quadratic phase functions [19] or low-correlating binary phase functions to the spectrum [20]. This is beneficial for CARS microscopy applications, allowing one to target a specific vibration and map a specific molecule within a sample.

For spectral focusing with binary phase shaping, the excitation spectrum is first divided into a pump and probe. The probe is selected as a small spectral slice, and the remaining supercontinuum pulse is then chosen as the pump. Using the pulse shaper to change the polarization of the probe by 90° relative to the pump decouples the probe from the pump. The pump produces a SB-CARS response along its polarization axis, acting as the pump, Stokes, and probe field. A polarizer is used to discriminate between the SB-CARS response of the pump spectrum and pump-probe spectrum. Rotating the polarizer to only allow the propagation of light along the polarization axis of the probe ensures that only the SB-CARS response of the probe is measured. This measurement has a greatly reduced non-resonant background due to the narrow bandwidth of the probe [15].

The pump part of the spectrum is shaped by further dividing the spectrum into two sections and adding two identical low-correlating binary sequences (with phase values $0$ and $\pi$) to the spectral phase of the two sections. The spectrum is divided to ensure that the start frequencies of the two sequences are separated by the spectral distance that matches the Raman frequency of interest. This process essentially creates spectral pairs within the pump that are spectrally separated from each other matching the chosen Raman frequency. These pairs are always in phase with each other but are poorly phase-correlated with other spectral components, thus establishing a preference to excite a chosen frequency.

## 3. Pulse characterization and compression

Pulse compression produces ultrashort pulses with high peak intensities required to drive nonlinear processes such as CARS. The supercontinuum produced by the ANDi-PCF is coherent and shot-to-shot stable, which enables pulse compression by the application of the inverse reconstructed phase from one measurement [21]. From a spectroscopic point of view, measurements are repeatable from pulse to pulse which means spectra can be averaged over many measurements to reduce noise.

The ANDi-PCF produces a supercontinuum with a spectral bandwidth that is ideal for performing SB-CARS spectroscopy that covers the fingerprint region [11]. Using an ANDi PCF does, however, present one with a challenge with regards to the spectral phase since the produced supercontinuum is stretched in time due to dispersion in the fiber. In order to use this supercontinuum as an excitation source for SB-CARS, the phase distortions from the generation process, propagation in the fiber, and contributions from other optical components must be compensated for.

Several methods exist to characterize ultrashort laser pulses, as they cannot be measured directly. The majority of these techniques, such as FROG [22] or SPIDER [23], are not convenient for characterizing pulses at the sample plane of a high numerical aperture microscope objective in a single-beam geometry. Hence, we implemented and compared two alternative pulse characterization/compression techniques that can easily be implemented into an existing SB-CARS setup with a spatial light modulator integrated in a 4f pulse shaper. We then determined the efficacy of each method in generating SB-CARS spectra using our particular light source. The first, multiphoton intrapulse interference phase scan (MIIPS) [9], is a popular femtosecond pulse compression technique, and the second, i$^2$PIE [10], is a more recent technique based on temporal ptychography developed in our laboratories and here implemented in a SB-CARS setup for the first time. Both of these techniques are phase only pulse characterization techniques which can be employed in the same experimental setup without the introduction of extra beamlines thus conserving the single-beam geometry of the SB-CARS setup.

### *3.1 MIIPS*

MIIPS has been the technique of choice in single beam microscopy applications as it delivers compressed pulses at the sample plane [9,24]. This is especially attractive for nonlinear spectroscopy and microscopy applications due to the dependence of nonlinear optical processes on the peak intensity of the excitation pulse. MIIPS exploits the relationship between the second harmonic generation (SHG) intensity spectrum produced through intrapulse interference and the spectral phase of the pulse being characterized.

Reconstruction of the spectral phase is facilitated by an approximation to the relationship such that the SHG intensity is dependent on the second order phase or, in other words, group delay dispersion (GDD). A maximum intensity is expected where the spectral phase is zero, which is exploited with the addition of a sinusoidal phase function that scans over the spectrum. A series of SHG intensity spectra are measured where the sinusoidal phase function overlaps with the spectrum at different spectral positions and constituted into a MIIPS trace.

The MIIPS trace is processed by mapping maximum intensities to phase values, to be reverse engineered into a first estimation of the spectral phase. Due to the approximation to second order spectral phase, this process must be repeated multiple times to determine an accurate spectral phase function, representative of the input pulse spectral phase. Comin et al. [25] provides a thorough and instructive explanation of the MIIPS algorithm and implementation.

### 3.2 i$^2$PIE

Recently time domain ptychography was introduced as a method for determining amplitude and phase of ultrashort pulses. Subsequently the time-domain ptychographic iterative engine was extended to generalized spectral phase-only transfer functions [10]. Hence similar to MIIPS, i$^2$PIE is a phase only pulse characterisation technique, capable of delivering optimally compressed pulses at the sample plane by determining the spectral phase from a set of SHG spectral intensity measurements. Implementations of time-domain ptychography rely on a fully characterised probe pulse in order to determine the amplitude and phase characteristics of an unknown pulse [26]. We extended the use of the ptychographic iterative engine (PIE) algorithm in order to reconstruct completely unknown object pulses with the application of arbitrary, known phase only transfer functions to the input pulse spectrum [10].

For our experiments, a set of quadratic phase functions were chosen and applied to the spectrum. The resulting SH spectra from a BBO crystal placed at the sample plane were measured. This set of second harmonic spectra were processed with the new i$^2$PIE algorithm and the spectral phase reconstructed. For this technique, only one measurement set is required to reconstruct the phase. This is in contrast to MIIPS, where a MIIPS trace is processed, the reconstructed phase is subtracted from the input pulse spectral phase, a new MIIPS trace is measured, and the process repeated multiple times until convergence is achieved.

### 4. Experimental setup and results

The experiments were performed using a single-beam line geometry with a Ti:Sapphire oscillator (Spectra Physics Tsunami) pumping a polarization maintaining all-normal dispersion photonic crystal fiber (NKT photonics NL-1050-PM-NEG) to provide broadband laser pulses. A chirped mirror (CM) pair is used for precompression of the pulses and corrects for the majority of chirp from the PCF as well as the dispersion of the optical components leading up to the sample plane (SPL). Phase modulation is performed using a pulse shaper constructed with 2 blazed gratings (600 l/mm), 2 cylindrical lenses (f = 250 mm) and a spatial light modulator (Jenoptik 640d) in a 4f geometry. A knife edge is placed into the Fourier plane of the pulse shaper on the short wavelength side of the spectrum to block the spectrum below 750 nm. The laser beam is focused onto the sample plane with a high numerical aperture microscope objective (60X Olympus Plan Fluorite Objective with Correction Collar) and collimated with another objective (40X Olympus Plan Fluorite Objective) before being focused into a USB-spectrometer (Avantes-uls3648). A shortpass filter (SP) is placed before the spectrometer to block the excitation pulse.

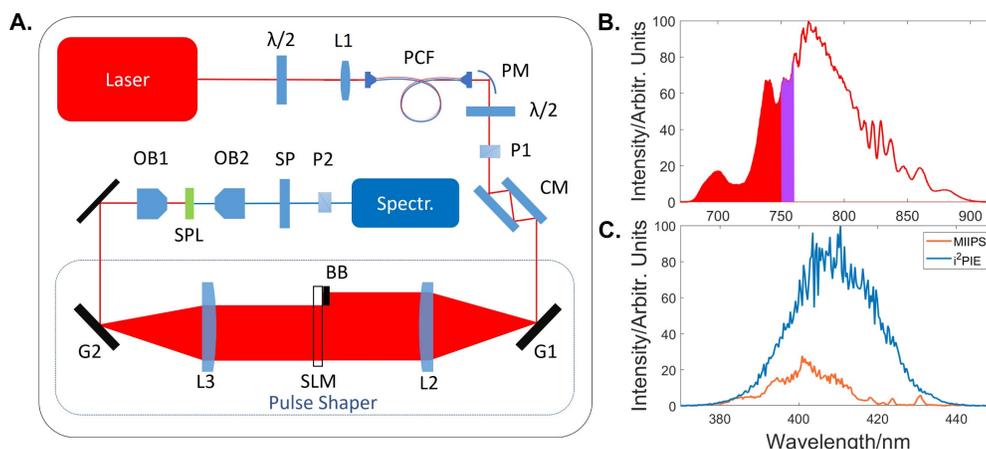

Fig. 1. The schematic representation of the SB-CARS setup (shown in **A**.), with the measured supercontinuum spectrum produced by the PCF in **B**. The red area in B shows the section of the spectrum that is removed by the knife's edge (BB), while the purple area corresponds to an exaggerated spectral slice chosen for probing. Compression of the supercontinuum using MIIPS (orange) and i²PIE (blue) yield the corresponding second harmonic spectra in **C.** The major components of the setup shown in A are halfwave plates (λ/2), lenses (L#), a photonic crystal fiber (PCF), polarizers (P#), a pair of chirped mirrors (CM), gratings (G#), a spatial light modulator (SLM), a knife edge (BB), microscope objectives (OB#), a shortpass filter (SP), and a spectrometer (Spectr.)

The same experimental setup is used to characterize and compress the supercontinuum pulses, as well as implement SB-CARS strategies. For pulse characterization measurements, the knife edge is removed and the shortpass filter is replaced with a bandpass filter (Thorlabs FGB39) that allows for the transmission of light between 360 nm and 580 nm. Phase functions pertaining to each characterization method are applied with in-house Labview software. Pulses are compressed by applying the inverse of the reconstructed phase with the pulse shaper to the supercontinuum spectrum. Due to the additive nature of spectral phase, one can simply add another phase function to the spectrum -such as a constant $\pm\pi/2$ phase over a small spectral region. Thus, one SLM in 4f-geometry is all that is required to fulfill phase manipulation criteria for both pulse characterization and SB-CARS.

The ANDi PCF, pumped with 650 mW average power (800 nm central wavelength, 80 MHz repetition rate, 80 fs pulse length and 8 nJ per pulse), produces a broad supercontinuum from 680 nm to 900 nm (figure 1.B). The full spectrum was characterized using the MIIPS and i²PIE techniques. Generating and measuring the second harmonic (SH) of the compressed pulse provides a measure of the degree of pulse compression. In figure 1.C the SH of the supercontinuum is shown for both characterization techniques. Both SH spectra are measured under the same experimental conditions and are directly comparable. Without direct characterization of the compressed supercontinuum, the SH generated spectrum allows us an indirect indication of how successfully the SC was compressed. Calculating the integrated SH intensity shows an increase by a factor of four, a clear indication that i²PIE results in shorter pulses after compression. The phase of a pulse spectrally broadened through self-phase modulation could not be adequately characterized using MIIPS [27], while i²PIE was better able to reconstruct the phase and thus yielded a shorter pulse after compression.

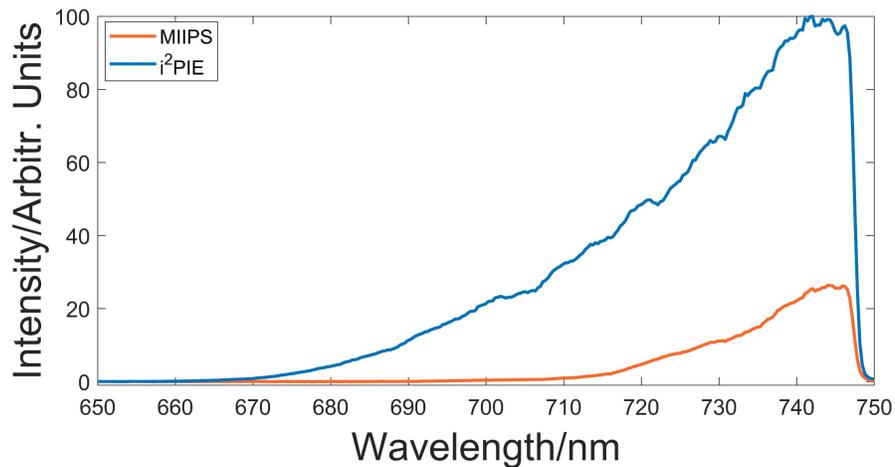

Fig. 2. Measured SB-CARS spectra with no phase shaping strategy applied apart from compression using i$^2$PIE (in blue) and MIIPS (in orange) algorithms.

A similar result can be observed when measuring the SB-CARS spectrum resulting from only the compressed pulses, i.e. the unprocessed superposition of both resonant and non-resonant signals. A liquid sample of para-xylene ($C_8H_{10}$) was placed in the sample plane and the resultant SB-CARS spectra produced by the supercontinuum compressed with MIIPS and i$^2$PIE was measured. In order to measure the SB-CARS spectra, a knife's edge was placed in the Fourier plane of the pulse shaper, which blocked out the spectral content with wavelengths below 750 nm (as indicated by the red area of the supercontinuum in figure 1.B). Due to the single-beam geometry and large bandwidth of the excitation source, the produced SB-CARS spectrum overlaps spatially and spectrally with the excitation spectrum. The addition of the knife-edge in the Fourier plane ensures that only the CARS signal is measured below 750 nm. A short pass filter (lambda < 750 nm) was placed after the sample to block the remaining supercontinuum, allowing only the induced CARS signal to reach the spectrometer.

Calculating the integrated intensity of both measurements show an increase by a factor of 6.5 (i$^2$PIE to MIIPS), slightly less than the factor of 8 that is expected when comparing to the factor 4 increase seen in the SH measurements. The measurement associated with the MIIPS compression approaches zero at around 700 nm while i$^2$PIE still shows a signal at 670 nm. This provides access to a larger probing range by using i$^2$PIE.

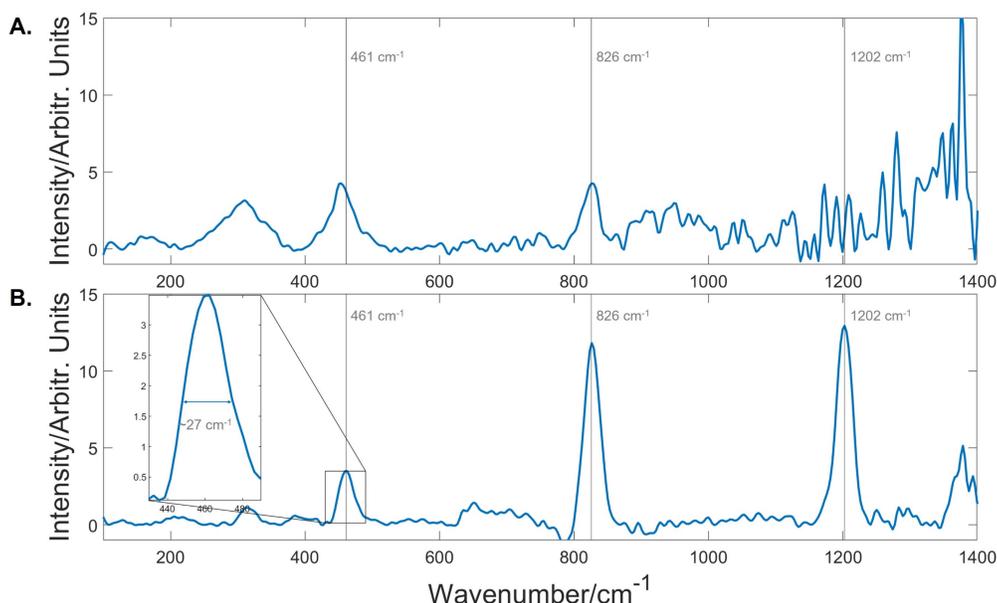

Fig. 3. Processed CARS spectra of para-xylene from SB-CARS measurements using the phase shaping of a small selection of the supercontinuum spectrum, using **A.** MIIPS and **B.** i$^2$PIE to compress the supercontinuum.

The CARS spectrum of para-xylene was extracted from the SB-CARS measurements by processing two measured spectra using equation 1 and is presented in figure 3. For these measurements two SLM pixels centred at 752 nm (corresponding to a bandwidth of ~22.4 cm$^{-1}$) were chosen to act as the phase probe. The probe is chosen to be in the high frequency part of the spectrum (as shown by the purple area in figure 1.B) to ensure that the response from the resonant Raman vibrations overlaps with the spectral region below 750 nm.

A visual comparison of the two results show a decreased probing range for the spectrum corresponding to MIIPS compression. This is evident in the presence of the 1200 cm$^{-1}$ peak in the measurement utilizing i$^2$PIE and the lack thereof in the measurement utilizing MIIPS. Without prior knowledge of the vibrational spectrum of para-xylene one would not be able to confidently differentiate between background and real peaks in the measurement utilizing MIIPS, but one can confidently identify peaks at 461 cm$^{-1}$, 826 cm$^{-1}$ and 1200 cm$^{-1}$ in the measurement utilizing i$^2$PIE. The highlighted peaks match those found in literature [28]. Analysis of the second spectrum, using the 461 cm$^{-1}$ peak, yields a spectral resolution of 26.9 cm$^{-1}$, which corresponds to the width of the spectral slice chosen as probe (~22.4 cm$^{-1}$).

The standard deviation of a selection of the background was calculated and chosen to represent a quantity of the background, which allows us to quantify the signal to background for our measurements. For the 826 cm$^{-1}$ the peak values were extracted for both measurements. The signal to background value obtained through this means for the MIIPS measurement is 16.6 and the value for i$^2$PIE is 65.1 an almost four-fold increase.

Implementation of the i$^2$PIE compression also allowed us to perform SB-CARS measurements using the spectral focussing technique previously discussed. This was not possbie using MIIPS compression as the response along the probe polarization is weak compared to measurements where the full spectrum acts as a probe. With the implementation of i$^2$PIE, however, we were able to measure an isolated peak.

For this measurement, liquid carbon disulfide was used as sample and the supercontinuum was compressed using the spectral phase reconstructed with i$^2$PIE. Two measurements were performed, one targeting a known carbon disulfide Raman line, 656 cm$^{-1}$, and one targeting a frequency that does not correspond to a Raman line in the carbon disulfide spectrum. A binary sequence with 30 bits was stretched over the identified regions and measurements were taken with a spectrometer integration time of 50 ms. An analyzer was added to the system in front of the spectrometer, to measure only light along the polarization of the probe.

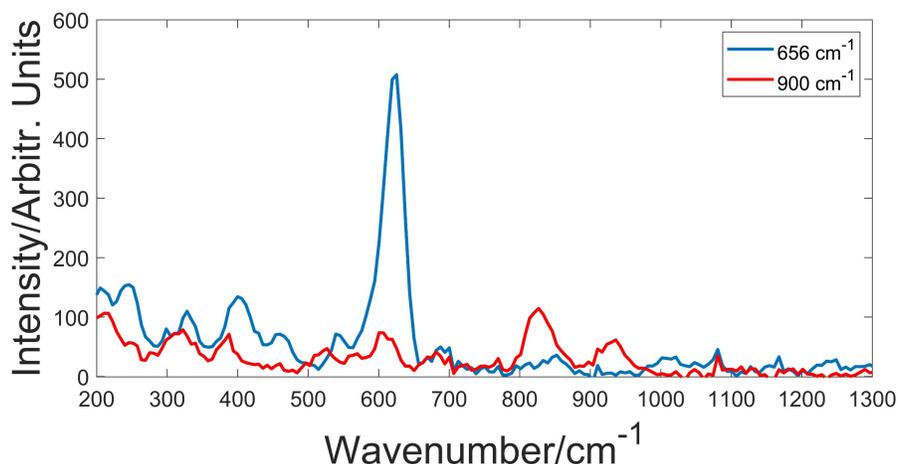

Fig. 4. SB-CARS measurements employing spectral focusing with binary phase shaping targeting a resonant Raman line of carbon disulfide at 656 cm$^{-1}$ (in blue) and targeting a frequency that is not a part of the carbon disulfide spectrum at 900 cm$^{-1}$ (in red).

Targeting the resonant Raman line of carbon disulfide successfully yielded a spectrum with the resonant Raman excitation isolated and surrounding background suppressed. As a test, the accompanying measurement (red line), shows that targeting a random frequency (in this case 900 cm$^{-1}$) suppresses the resonant Raman excitation. The response around 656 cm$^{-1}$ in the second measurement is indistinguishable from the surrounding background.

## 5. Conclusion

We have shown for the first time the implementation of i$^2$PIE for pulse characterization and compression with the aim of performing SB-CARS spectrometry. An ANDi-PCF was pumped using a commercial fs oscillator, providing high repetition rate broadband supercontinuum pulses, with extremely limited pulse to pulse fluctuations. This enabled the reproducible temporal compression of the fiber output using i$^2$PIE and MIIPS. A pulse shaper consisting of an SLM in a 4f geometry, allows for the implementation of these pulse characterization techniques and SB-CARS techniques in the same experimental setup. SB-CARS measurements demonstrated the improvement in compression achieved using i$^2$PIE over MIIPS for our laser source. The low pulse energies used, the stability of the setup, the high peak powers, together with the high repetition rate makes this laser source ideal for most forms of nonlinear microscopy. As proof, SB-CARS measurements of sample molecules using two different techniques, one measuring the complete CARS spectrum and the other targeting a specific Raman transition, were conducted. Results confirmed the excellent signal to background achieved with our setup and hence the possibility of using this system for SB-CARS microscopy.

## 6. Funding

Portions of this research was funded by the CSIR National Laser Centre (NLC-LREPA24-CON-001), the National Research Foundation (NRF), and Multi-modal imaging in biophotonics: Project No. PISA-15-FSP-002 Photonics Initiative of South Africa (PISA) of the Department of Science and Technology (DST). Schweizerischer Nationalfonds zur Förderung der Wissenschaftlichen Forschung (PCEFP2_181222, 200020-178812). Bursary assistance for Ruan Viljoen was provided by the South African Research Chair Initiative of the Department of Science and Innovation and National Research Foundation.

**7. Acknowledgement**

One of the authors (R. Viljoen) would like to thank Prof Mark Tame, Laser Research Institute, for continued bursary funding.

**8. Disclosures**

The authors declare that there are no conflicts of interest related to this article.